\documentclass[12pt]{iopart}
\usepackage{setstack}
\usepackage{iopams} 
\usepackage{upgreek}
\usepackage{graphicx}

\begin{document}

\title{Birefringence of small apertures for shaping ultrashort pulses}

\author{A M Nugrowati$^1$\footnote{Present address:
Quantum Optics Group, Huygens Laboratory, Leiden University, Niels Bohrweg 2, 2333 CA Leiden, NL}, S F Pereira$^1$ and
A S van de Nes$^2$\footnote{Present address:
ONE Simulations, Schipholweg 103, 2316 XC Leiden, NL}}
\address{$ö1$ Optics Research Group, Department of Imaging Science and Technology\\ Delft University of Technology, Lorentzweg 1, 2628 CJ Delft, NL}
\address{$ö2$ Department of Physics, Imperial College London, Prince Consort Road, London SW7 2BW, UK}
\ead{nugrowati@physics.leidenuniv.nl}

\begin{abstract}
For ultrashort pulses having different states of polarization, the experienced time delay when passing through small apertures is different. In the case of a small slit (or a circular aperture), we report a significantly stronger dispersion for the TE (or azimuthal) mode as compared to that for the TM (or radial) mode, creating a noticeable time delay between the two orthogonal polarization states, even for very thin apertures. The birefringent effect of small apertures is caused by waveguide mode dispersion. In essence, the propagation constant of the excited modes varies with wavelength differently for othogonal polarization states: it increases with the incoming wavelength for TE (or azimuthal) and remains constant for TM (or radial) mode. A fundamental understanding of this phenomenon helps to explain, for example, the use of small apertures as wave plates \cite{Chimento}. Furthermore, this effect can be exploited by tailoring the width and thickness of the aperture to obtain the desired pulse-shape and delay. 
\end{abstract}

\pacs{42.25.Ja, 42.25.Lc, 42.65.Re, 42.79.Ag}

\maketitle 

\section{Introduction}
Advances in pulsed-laser physics have recently resulted in pulse lengths down to the femtosecond scale, which corresponds in the optical regime to only a few cycles of oscillations \cite{Keller}. These ultrashort pulses offer an even greater accuracy and sensitivity in a whole range of experiments, such as micro-machining, femtochemistry, multi-photon fluorescence microscopy, terahertz generation and many other fields, improving the understanding of the fundamental phenomena studied. The interaction of a pulse with any type of optical component is bound to introduce dispersion and therefore changes, normally broadens, the pulse-shape. A good understanding of the dispersion effects by the optical elements commonly used in an experimental set-up is therefore essential. 

Of a fundamental importance is the small aperture of finite thickness. Apertures with simple geometries have been studied in great detail, yielding advances in for example diffraction theory \cite{Bouwkamp}, and more recently advances in near field optics, e.g. extraordinary transmission and excitation of surface plasmons \cite{Ebbesen}. Earlier studies emphasize on the interaction between small aperture and light with a transverse magnetic (TM) mode, which supports the presence of surface waves. Additionally, a waveguide model for a perfect conductor forbids the presence of transmitted light with a transverse electric (TE) mode for apertures having widths below the cut-off dimension of $\lambda/2$. In his seminal paper \cite{waveguideslit}, Schouten \textit{et al.} showed that in the optical regime, the finite conductivity of the materials surrounding the aperture allows for extraordinary transmission, even below the cut-off width for TE-polarized light.   Experimentally, it has been recently demonstrated that TE-polarized light is still present after transmission through a very small slit, and in fact plays a role when tailoring a slit as a quarter wave plate \cite{Chimento}. Previous studies have shown that diffraction and scattering of a light pulse by a geometry of dimensions close to the wavelength in the optical regime result in a field distribution that is strongly polarization-dependent \cite{AMN, Mittleman}.

In this article, we discuss the birefringent effect of small apertures when illuminated by an  ultrashort pulse. We study the illumination of (i) a 1-D slit structure, as well as (ii) a 2-D circular aperture with an ultrashort pulse. Even though commonly used apertures are typically very thin, the encountered dispersion effects are significant due to the small width of the aperture in combination with the broad bandwidth of the pulse. In \Sref{results} we demonstrate that the dispersion effect by an aperture is strongly polarization-dependent. We report that the incoming pulse with field oscillating parallelly to the aperture experiences a larger time delay compared to that oscillating perpendicularly. To elucidate the strong polarization-dependent dispersion behaviour, we present the spectral distribution of the excited waveguide modes and the contributing propagation constants in \Sref{discuss}. This fundamental knowledge can be exploited for several possible applications such as pulse shaping, optical switching, or creating optical retarders using small apertures.

\section{Optical configuration}
\subsection{Illumination, material and geometry}
\begin{figure}[htbp]
\begin{center}
\includegraphics[width=12cm]{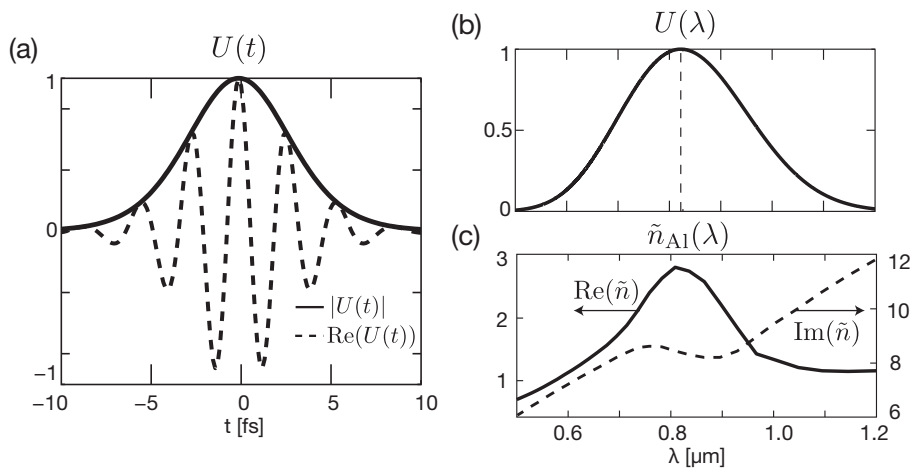}
\end{center}
\caption{\label{Fig01}(a) The time trace of an ultrashort pulse consisting of five cycles within a time span of $7.5~\rm{fs}$ and (b) its corresponding spectral distribution, with the peak at $825~\rm{nm}$. (c) The dispersion relation of aluminum (\~n$_{\mathrm{Al}}$) has a resonance frequency that coincides with the corresponding central wavelength.}
\end{figure}

A realistic model of an ultrashort pulse in the optical regime is based on an experimentally generated femtosecond pulse \cite{Poppe}. To avoid non-causal artifacts in the time response which are present in a Gaussian model \cite{Sheppard}, we have used only positive frequencies as described in \cite{Lefrancois}. To match the experimental conditions, we have chosen $\omega_0 = 1.2 \times 10^{15}~\rm{rad/s}$ the cutoff frequency, $\omega_1 = 1.2 \times 10^{15}~\rm{rad/s}$ which is approximately equal to the spectral bandwidth of the pulse, and $s = 10$ a positive integer that is proportional to the number of cycles in the pulse. Based on these parameters, the ultrashort pulse contains five cycles, has a total length of $7.5~\rm{fs}$, and a bandwidth of $500<\lambda<1200~\rm{nm}$, with its peak at $\lambda_{0}=825~\rm{nm}$. We show in \Fref{Fig01}(a) the time-trace $U(t)$, and in \Fref{Fig01}(b) the corresponding wavelength distribution $U(\lambda)$ of the ultrashort pulse. 

\begin{figure}[htbp]
\begin{center}
\includegraphics[width=8cm]{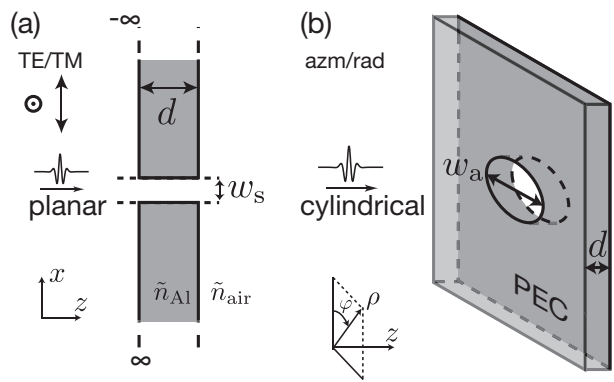}
\end{center}
\caption{\label{Fig02}Geometry of the apertures.  (a) The slit aperture is embedded in an aluminum layer, while (b) the circular aperture is embedded in a perfectly electric conducting (PEC) layer. We assume that the structures are illuminated by an ultrashort pulse at normal incidence with a planar wave for the 1-D slit structure case, and a cylindrically symmetric distribution of a converging wave for the 2-D circular aperture case.}
\end{figure}

In the case of a 1-D slit structure, the aperture is embedded in aluminum that has a dynamic dispersion relation \~{n}$_{\mathrm{Al}}$($\lambda$) in the optical region of interest, shown in \Fref{Fig01}(c). Also, to demonstrate the general validity, we extend the problem into a three-dimensional case where a 2-D circular aperture is embedded in a perfectly electric conducting (PEC) layer. To take full advantage of the symmetry of the geometry, we separate the field in a transverse electric and a transverse magnetic component, corresponding to linear polarization (TE/TM) states for the slit and cylindrical polarization (azimuthal/radial) states for the circular aperture. The slit aperture is schematically shown in \Fref{Fig02}(a), and the circular aperture in \Fref{Fig02}(b). The aperture dimensions are chosen to be near, but smaller than the center wavelength of the pulse. It is $d=700~\rm{nm}$ thick and has a width of $w_{\mathrm{s}}=700~\rm{nm}$ for the slit, and has a diameter of $w_{\mathrm{a}}=1500~\rm{nm}$ for the circular aperture. 

\subsection{Method of calculation}
We have used the modal decomposition technique in our calculation, since it allows us to express the field inside the aperture by a set of coefficients corresponding to the waveguide modes of the aperture. Each of these modes has a characteristic field distribution in the transversal plane, and a specific propagation constant which is directly related to dispersion. The coupling coefficients indicate the amount of energy available in each mode and are determined by matching the waveguide modes with the field outside the aperture using the boundary conditions. 

In this paper, we apply two different rigorous vectorial methods --- namely (i) the Fourier modal method for the 1-D slit case; and (ii) the modal method \cite{Roberts} for the 2-D circular aperture case. Details of the method to calculate the propagation of an ultrashort pulse through a 1-D slit can be found in our previous work \cite{AMN}, where we have assumed an incoming plane with a temporal evolution as described in \Fref{Fig01}(a). For the case of a 2-D circular aperture in \Fref{Fig02}(b), the modal method is best suited for modelling problems in PEC materials since analytic expressions of the field expansion are known. Mathematically, the incoming beam with cylindrically polarized state using cylindrical coordinates ($\rho,\varphi,z$) can be expressed by a Bessel-Gauss beam, defined by the function:
\begin{equation}\label{eq:circular-illumination}
U(\mathbf{r})=w_{\mathrm{p}} k_0 \int_0^{k_0}{\frac{k_0 k_{\rho}}{k_z^2}J_1(\rho k_{\rho})e^{-\left(\frac{w_{\mathrm{p}} k_0 k_{\rho}}{2 k_z}\right)^2}e^{i k_z z}\mathrm{d}k_{\rho}}\;,
\end{equation}
with $U$ either the azimuthally polarized electric or magnetic field in the unit $[\mathrm{V/m}]$ or $[\mathrm{A/m}]$, respectively; $w_{\mathrm{p}} = 1~\upmu\mathrm{m}$ is proportional to the transverse beam extent of the pulse; $k_0$ is the wave vector in vacuum and a function of wavelength, with $k_{\rho}$ and $k_z$ the wave vector components in cylindrical coordinates. Because of the choice in polarization state of the pulse, the field distribution is described by the first order Bessel function of the first kind $J_1$.

\section{Results}\label{results}
Changes in the properties or shape of the pulse during propagation, for instance envelope broadening and pulse chirping, can be attributed to dispersion effects. Dispersion is, in essence, an effect caused by a frequency dependence of the phase velocity $v_\mathrm{p}$ (zeroth-order dispersion) and/or group velocity $v_\mathrm{g}$ (first-order dispersion). In a waveguide structure, dispersion is also defined as the dependence of the phase velocity $v_\mathrm{p}$ in a medium on the propagation mode \cite{Snyder}. 
\begin{figure}[htbp]
\begin{center}
\includegraphics[width=13cm]{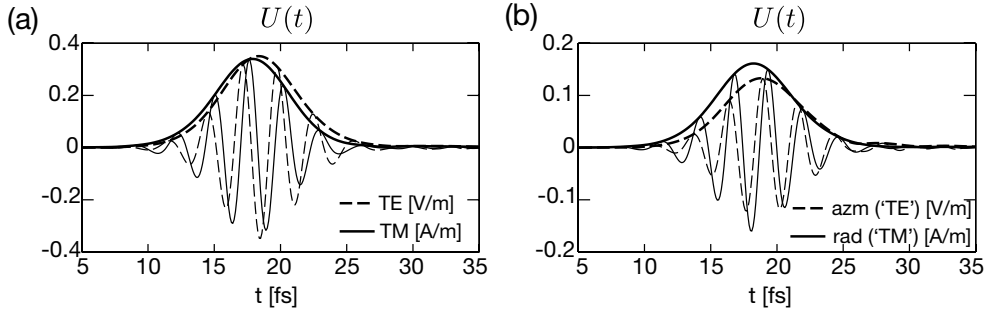}
\end{center}
\caption{\label{Fig03}Transverse electric (dashed) and magnetic (solid) component at (a) $z=5~\upmu\rm{m}$ from the centre of the slit aperture, and at (b) $z=5~\upmu\rm{m}$ and $\rho=2~\upmu\rm{m}$ from the center of the circular aperture, illuminated by an ultrashort pulse.}
\end{figure}

As shown in \Fref{Fig03}, a metal slab with a small aperture induces dispersion. The pulse after transmission through the apertures is shown in \Fref{Fig03}(a) for the 1-D slit and \Fref{Fig03}(b) for the 2-D circular aperture, as observed at a distance of $z=5~\upmu\rm{m}$ from the center of the aperture, located on the optic axis for the slit and $2~\upmu\rm{m}$ off the optic axis for the circular aperture. The phase velocity $v_\mathrm{p}$ strongly affects the shape of the pulse oscillation, i.e the real field of the pulse. The shape of the pulse envelope on the other hand is affected by the group velocity $v_\mathrm{g}$, which is the amplitude of the field. The real field that also expresses the spectral contents of the pulse is usually the measured quantity in the experiments.
%

A comparison of the transverse electric and the transverse magnetic field component shows a difference in the optical path length of the pulse envelope equivalent to approximately $0.5~\rm{fs}$ (about 7\% of the pulse length). The TM-component arrives earlier then the TE-component, meaning it experiences less group delay (or smaller first-order dispersion). For both structures under study, we observe a small second-order dispersion, a slightly broadened pulse envelope.

Note that similar dispersion effects have been observed by propagation of much slower pulses through kilometers long fibers, where minute dispersion effects are still of concern due to the very long path length \cite{Poole}. Although the pulse passing through these structures experiences dispersion effects that are dominated by the first-order dispersion, this dispersion effect is still of importance since it influences the spectral content which is an important information in measurements. To explain the strength and difference in dispersion for an aperture of only $700~\rm{nm}$ thickness, we study the propagation constant of the waveguide modes contributing to the field inside each aperture. This study demonstrates the \textit{waveguide dispersion} that includes several modes, and therefore also involves the \textit{intermodal dispersion} effect, i.e. the group delay difference between the different modes \cite{Snyder}. 

\section{Discussions}\label{discuss}
\subsection{Dispersion by a small slit}
\begin{figure}[htbp]
\begin{center}
\includegraphics[width=15cm]{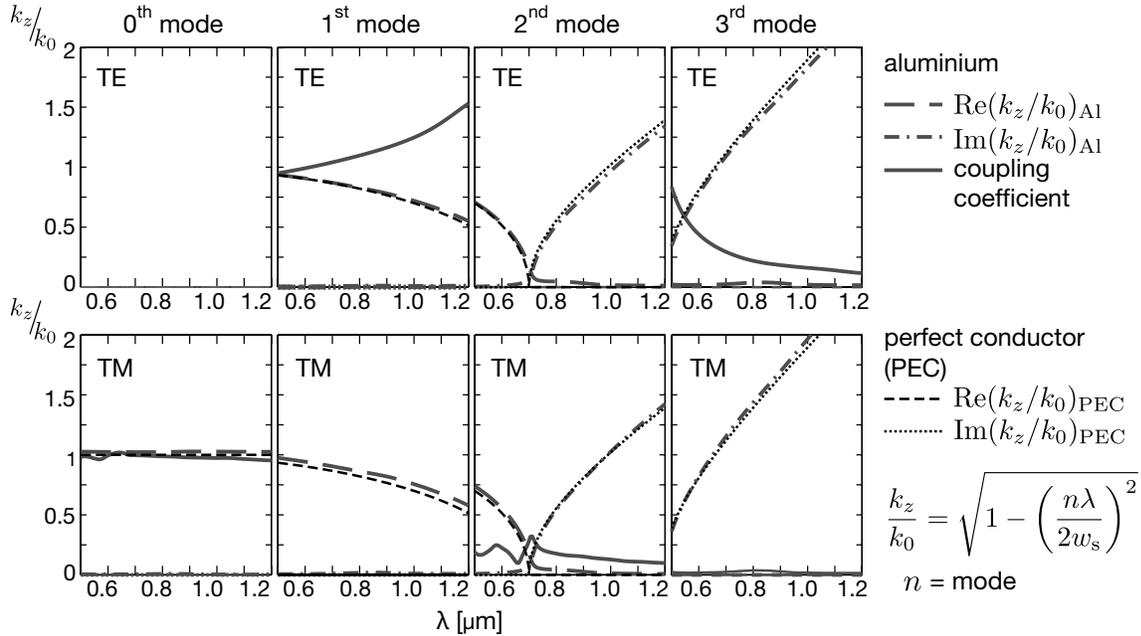}
\end{center}
\caption{\label{Fig04} The real (long-dashed lines)  and imaginary part (dash-dotted lines) of the modal propagation constant ${k_z/k_0}_{\mathrm{Al}}$ for an aluminum slit, as well as the corresponding absolute value of the coupling coefficient (solid lines) as a function of the wavelength. For a comparison, we also show the real (short-dashed lines) and imaginary (dotted lines) part of the propagation constant ${k_z/k_0}_{\mathrm{PEC}}$ for a slit in a PEC layer.}
\end{figure}
The waveguide modes of the aperture form a complete orthonormal set of solutions, and a linear combination of these modes can be used to fully represent any field inside the waveguide. For a slit in aluminum, these modes will be similar but not identical to the  sine (TE) and cosine (TM) functions with the argument $n \pi x/w_{\mathrm{s}}$ where $n$ is the mode number. The linear combination of the modes have been solved numerically using the Fourier modal method as discussed in \cite{Moharam}, allowing for non-linear coordinate transformations \cite{Hugonin}. In the limiting case of PEC, we obtain as an exact solution the sine and cosine modes. In \Fref{Fig04}, we have plotted the complex propagation constants and the absolute value of the coupling coefficients as a function of the wavelength for the four lowest energy waveguide modes. Note that the value of the propagation constant has been normalized on $k_0$, the wave number, which is of course a function of the wavelength as well. In the case of the perfect electric conductor, the normalized propagation constant is $k_z/k_0=\sqrt{1-\left(n\lambda/2w_{\mathrm{s}}\right)^2}$
with $n$ the mode number. All the coupling coefficients have been normalized on the energy of the pulse illuminating the aperture area. 

The plane wave illumination couples only to the modes with an odd mode number for TE illumination and only to modes with an even mode number for TM illumination. As expected from the boundary conditions for the transverse electric field, the zeroth mode does not exist. The real component of the propagation constant for propagating modes is inversely proportional to the phase velocity $v_\mathrm{p} = c (k_0/k_z)$. The phase velocity of each waveguide mode is always equal to or larger than the speed of light $c$. However the propagation speed of the pulse is determined by not only the longitudinal but also the transversal propagation constant of the mode, and can be approximated by 
\begin{equation}
v_\mathrm{g} = c \left[(k_z/k_0)-\lambda \frac{\rm{d}(k_z/k_0)}{\rm{d}\lambda}\right]^{-1}.
\end{equation}
The group velocity of each waveguide mode is always equal to or smaller than the speed of light, and for the case of PEC, it simplifies to $v_\mathrm{g}=c(k_z/k_0)$.

For the TE illumination, the main contribution is given by the first mode. Since the real part of the normalized propagation constant ($k_z/k_0$) of this mode is smaller than unity and monotonically decreasing over the entire wavelength range, the pulse experiences the first-order dispersion effect, i.e. the frequency dependence of the phase delay. In order to reconstruct the pulse, each mode needs to be multiplied with its coupling coefficient, representing the excitation strength. The variation in amplitude of the coupling coefficients as a function of the wavelength has only a minor influence on the actual pulse envelope (i.e. the amplitude of the field). However, the oscillation profile of the pulse (i.e. the real field of the pulse) is considerably changed due to the phase of the coupling coefficients in combination with the propagation constant of the mode that vary as a function of the wavelength. The contribution of the third mode strongly favors the smaller wavelengths ($w_{\mathrm{s}}>\lambda/2$) but is only observable for very thin apertures, since the dominant imaginary part of the propagation constant results in a rapid decay of that mode. 
%

For the TM illumination, the main contribution is given by the zeroth mode. Notice the uniform wavelength dependence of the modal coefficient and the propagation constant. This explains the pulse envelope that is almost identical to that propagating in free-space, i.e. no dispersion effects are observed for the zeroth mode. The contribution of the second mode is restricted to wavelengths up to $\lambda = 700~\mathrm{nm}$, however this contribution experiences a significant delay as compared to the zeroth mode due to the small and decreasing real part of the propagation constant. For larger wavelengths, the mode is evanescent, and the frequency dependence of the group delay increases.

\subsection{Dispersion by a small hole}
\begin{figure}[htbp]
\begin{center}
\includegraphics[width=15cm]{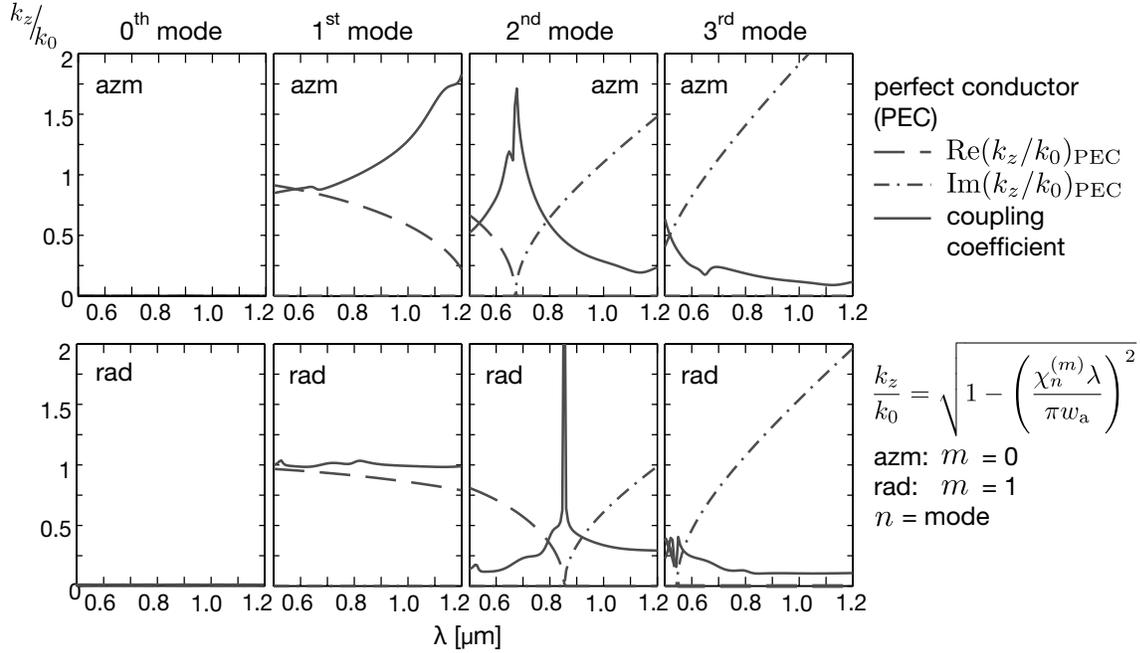}
\end{center}
\caption{\label{Fig05}The real (long-dashed lines) and imaginary part (dash-dotted lines) of the modal propagation constant $k_z/k_0$, as well as the absolute value of the coupling coefficient (solid lines) as a function of the wavelength.}
\end{figure}
A similar analysis can be applied for the case of a 2-D circular aperture in a perfect electric conductor, using the formalism described in \cite{Roberts} for the modal method. Since we apply azimuthally uniform, radially- or azimuthally-polarized illumination, only a subset of the cylindrical waveguide modes can be excited. The relevant amplitude functions of the waveguide modes are given by the Bessel functions $J_1(2\chi^{(1)}_n \rho/w)$ and $J_1(2\chi^{(0)}_n \rho/w)$ for azimuthally (corresponding to TE) and radially (corresponding to TM) polarized illumination, respectively, with index $n$ as the mode number. The constants $\chi^{(m)}_n$ are the roots of the Bessel function of the first kind defined by $J_m(\chi^{(m)}_n)=0$, with index $m$ as the order of the Bessel function, and index $n$ as the root index. 

In \Fref{Fig05}, we have plotted the complex propagation constants and the absolute value of the coupling coefficients as a function of the wavelength for the four lowest energy waveguide modes. Again, the coupling coefficients have been normalized to the energy of the pulse illuminating the aperture area, and the propagation constants have been normalized on $k_0$ given by the following expression
\begin{equation}
\frac{k_z}{k_0} = \sqrt{1-\left(\frac{{\chi_{n}^{(m)}}\lambda}{\pi w_{\mathrm{a}}}\right)^2},
\end{equation}
with $m=0$ for the azimuthal and $m=1$ for the radial polarization, and $n$ the mode number. In contrast to the 1-D slit case, both polarization states can now excite all modes, except the zeroth. This is required by the zero field along the optical axis of all modes due to the cylindrical nature of the polarization. 

For both the azimuthally and the radially polarized illumination, the main contribution is given by the first mode. However, for the azimuthally polarized illumination, the propagation constant decreases significantly as a function of the wavelength, while for the radially polarized illumination the dependence is more uniform. The contribution of the second mode depends strongly on the wavelength, since the propagation constant becomes evanescent from $\lambda=672~\rm{nm}$ and $\lambda=854~\rm{nm}$, for azimuthally and radially polarized illumination, respectively. The different cutoff wavelengths for both polarizations states are a result from the difference in mode distribution and corresponding propagation constant. Due to the PEC nature of the material, the propagation constant changes from purely real to purely imaginary at the cutoff wavelength. For the third mode, only a small wavelength range up to $\lambda=545~\mathrm{nm}$ for radially polarized illumination, contributes to the transmitted field. The net dispersion effect of a circular aperture is also dominated by the first-order dispersion. The radially polarized pulse experiences a smaller first-order dispersion effect, with a smaller group delay as compared to the azimuthally polarized pulse. Note that, however, the oscillation profile of the radially polarized pulse is changed as compared to the incoming pulse, indicating the contribution of the strong dispersion effect at the higher order modes.

\subsection{Role of the aperture thickness}\label{thickess}
\begin{figure}[htbp]
\begin{center}
\includegraphics[width=12cm]{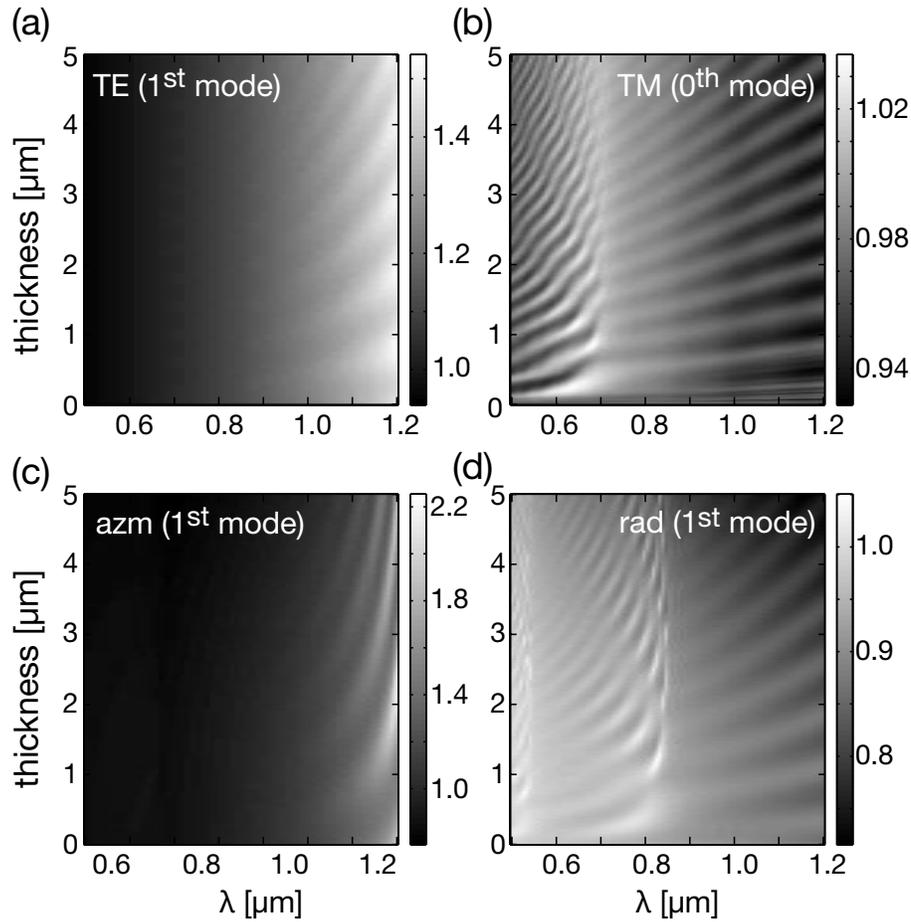}
\end{center}
\caption{\label{Fig06}Absolute value of the coefficient as a function of the wavelength and the aperture thickness, with (top) the $1^{\rm{st}}$ and $0^{\rm{th}}$ mode for (a) TE and (b) TM illumination of the slit, and (bottom) the $1^{\rm{st}}$ mode for (c) azimuthally and (d) radially polarized illumination of the circular aperture, respectively.}
\end{figure}
Another effect to consider in order to understand the exact oscillation profile of a pulse, is the dependence of the modal contribution at specific wavelengths on the exact thickness of the aperture. Interference effects influence the transmission, which shows up as oscillations of the the absolute value of the coupling coefficients as a function of the thickness of the aperture and the wavelength, shown in \Fref{Fig06}. As only a single mode is excited at larger wavelengths, the oscillations of the amplitude follow closely the standard Fabry-P\'{e}rot interferometry relations with minima when $d = N \pi/k_z$ where $N$ denotes an integer. The boundary at which the single mode excitation and a standard Fabry-P\'{e}rot behaviour start to occur, can be seen more clearly in the TM (right) illumination cases. 

The choice to consider only azimuthally or radially polarized illumination for the circular aperture, was made from an instructional point of view such that the excited modes uncouple. For different types of illumination other coupled modes will be excited what complicates the analysis, but similar dispersion effects between the TE and TM response are present since these effects are determined by the waveguide shape and dimensions.

\section{Potential applications}
In general, the transmitted TE/azimuthal or TM/radial pulse exhibit \textit{birefringent retardation}. One can therefore tune the geometry of such small apertures to create wave plates, as demonstrated in \cite{Chimento}. It is also important to realize that similar dispersion effects can be observed in case of arrays of apertures as well. Furthermore, the exact form of response can be tuned by a careful selection of the size and thickness of the apertures, leading to a whole range of potential applications. For example, it is possible to use the observed polarization-dependent dispersion effects to our advantage by an intelligent use of the delay between both polarization components in pump-probe experiments. A desired $2~\rm{fs}$ delay between both polarization states for a slit of $700~\rm{nm}$ width would correspond to $d=2~\upmu\rm{m}$ thickness. Also, the shape of the transmitted pulse for both polarization states can be modified in such a way to trigger or observe different types of effects. Finally, the dispersion effects can be of particular concern for the interpretation of near field measurements using either SNOM-tips or small apertures and should therefore be considered during the analysis.

\section{Conclusion}

In conclusion, we have shown the importance of strong polarization-dependent dispersion effects for the transmitted ultrashort pulse passing through thin and small apertures, indicating the birefringence property of such apertures. These effects will become increasingly important due to the recent advances in availability of these pulses, as well as the increasing interest in near-field analysis. 

Generally, the TM-polarized light is considered to have more interesting properties as compared to the TE-polarized light. In this study, we found that the polarization-dependent dispersion effect of the transmitted pulse oscillating along the TE direction is significantly larger than for the TM pulse, resulting in a noticeable group delay between both polarization states even for very thin apertures. As TM-polarized light is associated with surface waves, in a similar manner TE-polarized light should be associated with increased dispersion effects. These dispersion effects are caused by the difference in wavelength dependent propagation constants of the excited modes, related to the \textit{waveguide dispersion}, and the \textit{intermodal dispersion} effects. 

Whilst the transmission through a slit is dominated by the $0^{\mathrm{th}}$ (TM) and the $1^{\mathrm{st}}$ (TE) mode, and corresponding difference in first-order dispersion, the transmission through a circular aperture is dominated by the $1^{\mathrm{st}}$ mode for both the radial (TM) and azimuthal (TE) polarization. However, even for the circular aperture, still a noticeable difference in first-order dispersion exists.

The dispersion effects can be exploited by tailoring the width and thickness of the aperture to shape the envelope and the oscillation profile of the pulse, as well as the delay between two orthogonally polarized states. Tuning the geometry of the aperture allows a match of the pulse envelope and delay to the requirements of pump-probe experiments, or can be used for creating variable pulse retarders. Cavity resonances inside the aperture should be considered while fine-tuning the contribution of particular wavelength components. Finally, including the dispersion effects of the aperture, in combination with the diffraction effects \cite{AMN}, is essential for a correct analysis of the results from optical experiments using ultrashort pulses.


\end{document}